\documentclass[pdflatexm,sn-aps
]{sn-jnl}

\usepackage{graphicx}
\usepackage{multirow}
\usepackage{amsmath,amssymb,amsfonts}
\usepackage{amsthm}
\usepackage{upgreek}
\usepackage{mathrsfs}
\usepackage[title]{appendix}
\usepackage{xcolor}
\usepackage{textcomp}
\usepackage{manyfoot}
\usepackage{booktabs}
\usepackage{algorithm}
\usepackage{algorithmicx}
\usepackage{algpseudocode}
\usepackage{listings}
\usepackage{xcolor}

\newcommand{\com}[1]{#1}

\begin{document}

\title[]{Simple tunable phase-locked lasers for quantum~technologies}

\author[1]{\fnm{Nicola} \sur{Agnew}}\email{nicola.agnew@strath.ac.uk}
\equalcont{These authors contributed equally to this work.}
\author[1]{\fnm{David} \sur{Lowit}}\email{david.lowit@durham.ac.uk}
\equalcont{These authors contributed equally to this work.}
\author*[1]{\fnm{Aidan S.} \sur{Arnold}}\email{aidan.arnold@strath.ac.uk}

\affil[1]{\orgdiv{SUPA Department of Physics}, \orgname{University of Strathclyde}, \orgaddress{\street{107~Rottenrow}, 
\city{Glasgow}, \postcode{G4 0NG}, 
\country{UK}}}

\abstract{In a wide range of quantum technology applications, ranging from atomic clocks to the creation of ultracold samples for atom interferometry, optimal laser sources are critical. In particular, two phase-locked laser sources with a precise difference frequency are needed for efficient coherent population trapping (CPT) clocks, gray molasses laser cooling, or driving Raman transitions. Here we show how a simple cost-effective laser diode can selectively amplify only one sideband of a  fiber-electrooptically-modulated seed laser to produce moderate-power phase-locked light with sub-Hz relative linewidth and tunable difference frequencies up to $\approx 15\,$GHz. The architecture is readily scalable to multiple phase-locked lasers and could conceivably be used for future on-chip compact 
laser systems for quantum technologies.}


\maketitle

\section{Introduction and motivation}\label{sec1}

Quantum technology is burgeoning, and there are a wide variety of application areas requiring laser light with a frequency spectrum comprising exactly two modes that are phase-locked to each other -- essentially frequency-offset laser `clones' ideally with a delta-function frequency beat note. This paper demonstrates a simple two-laser system fulfilling this criterion, with each laser power $>  100\,$mW at its  frequency. This is relevant for applications in both thermal and ultracold CPT atomic clocks \cite{Knappe2004,AbdelHafiz2017,Liu2017,Elgin2019,Elvin2019}, Raman pulses for mirrors and beamsplitters in atom interferometry \cite{Dickerson2013,Rakholia2014,Wu2017,Morel2020,Abend2023}, as well as pulses relevant for logic gates in quantum computing with both atoms and ions \cite{Blatt2012,McDonnell2022,Graham2022,Evered2023,Bornet2023}, and Doppler-broadening thermometry \cite{Truong2011,Gravina2024,Agnew2026,Bresler2026}. 
We foresee utility in all of these areas, but also give more detail in Section~\ref{GM} for the specific moderate-power application of sub-Doppler laser cooling using the topical technique of gray molasses (GM) \cite{Boiron1995}. 

Regardless of atomic species, a crucial requirement for \com{many} of the above quantum technology applications is phase coherence between two precisely controlled laser frequencies incident on the atoms \cite{Rosi2018}.  
Our laser system is capable of cost-effectively producing the requisite moderate-power dual-frequency phase-coherent light, with a widely tunable difference frequency. The system is also designed to be integrated into existing laser cooling systems. 

Electro-optic modulators (EOMs \cite{Kawanishi2007,Kawanishi2022}) are ideal for making multi-frequency phase-coherent light, without independent laser systems. However, while free-space EOMs can handle higher optical input powers (Watts), they have low fractional sideband power -- particularly for Rb and Cs, with their large ground state hyperfine splittings -- with only $0.1\%$ relative RF frequency tunability due to the necessary resonant RF drive circuit \cite{newfocus}. Conversely, fiber EOMs have extremely wide frequency tunability, and large fractional sidebands for low RF input power -- but at wavelengths suitable for most alkali metal transitions they can only carry low optical powers ($\approx 25\,$mW) without damage, with $\approx 4\,$dB insertion loss \cite{iXblue}. 

A major disadvantage of both fiber and free-space EOMs is that any positive frequency sideband has an equal amplitude negative frequency sideband, in addition to the carrier. There are therefore always unwanted laser frequency sidebands -- which in the best case are wasted laser power, and in the worst case cause resonant heating or lead to light shifts affecting the performance of e.g.\ atomic clocks. IQ modulators \cite{Li2017,Zhu2018}, Serrodyne \cite{Johnson2010} and other alternative techniques \cite{Macrae2021,Dammalapati2025} offer the ability to make approximately single-sideband modulation, but without the cost-effectiveness, simplicity or flexibility we demonstrate here. We also suspect that our technique is likely to suppress EOM-induced residual amplitude modulation \cite{Gillot2022}.

\section{The phase-locked laser system}

\begin{figure}[!t]
    \centering
    \includegraphics[width = .95\columnwidth]{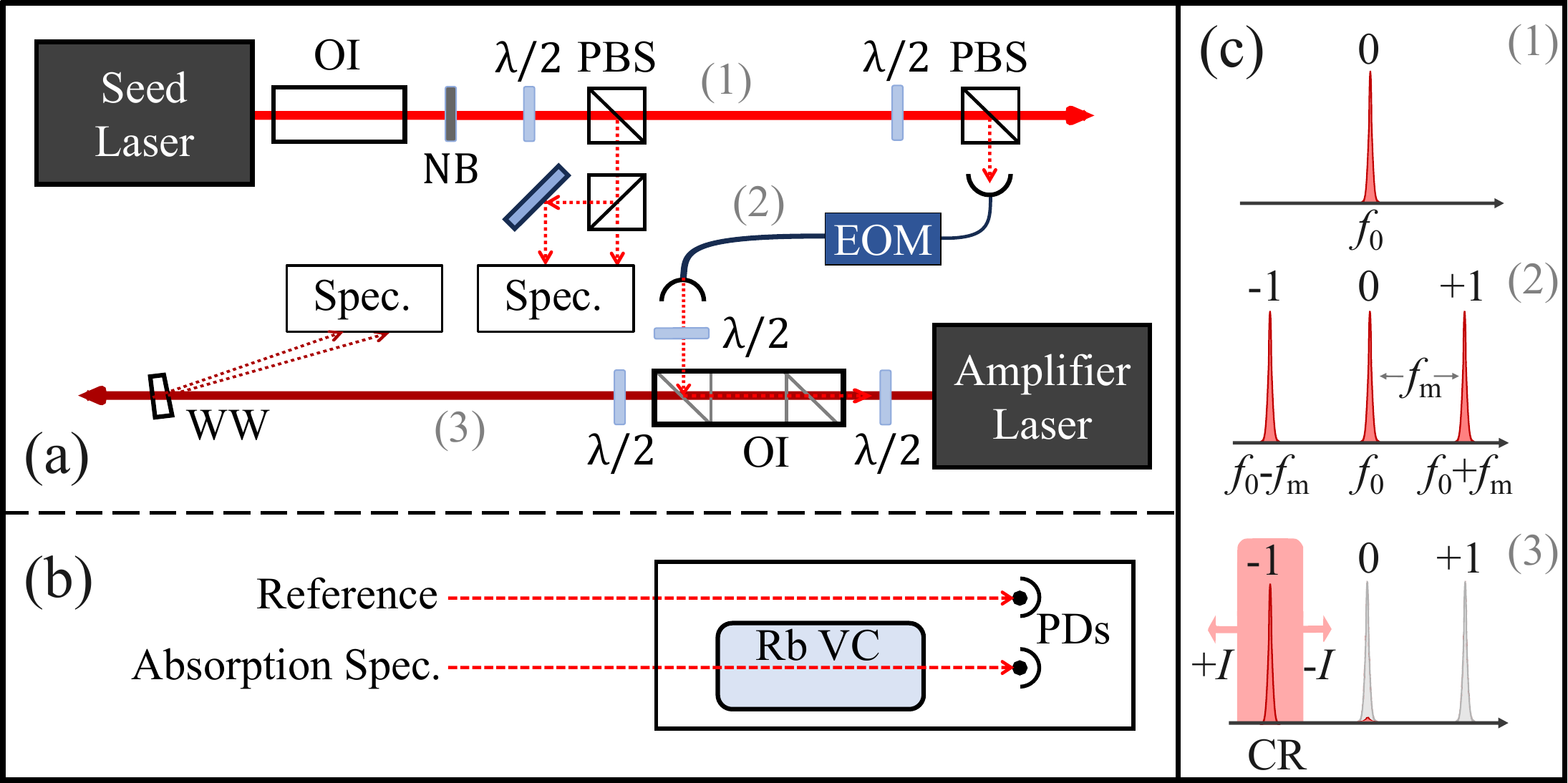}    
    \caption{Experimental diagram (a), with the seed and amplifier laser (SL, AL) low-intensity spectroscopy setup in (b). Abbreviations: optical isolator (OI), narrow-band frequency filter (NB), half-wave plate ($\lambda/2$), fiber electro-optic modulator (EOM), polarising beamsplitter (PBS), wedged window (WW), photodiode (PD), and vapor cell (VC). \com{(c) Conceptual diagram of the 3 key stages of sideband injection locking. The SL runs at a single frequency ($f_0$) in stage (1). In stage (2) the light passes through a fibre-coupled EOM generating sidebands at integer multiples of the modulation frequency ($f_m$). In stage (3) the modulated light is injected into the AL diode. The AL selectively amplifies only sidebands that fall within the capture range (CR), with a center frequency which can be selectively tuned by adjusting the AL diode current ($I$), acting as a diode length control.}}
    \label{fig:diag}
\end{figure}

A key part of our phase-locked laser system (Fig.~\ref{fig:diag}) is optical injection locking (OIL), whereby a small fraction of light $(<1\%)$ diverted from a seed laser (SL \cite{Kakarla:18}) is injected into a temperature-stabilised amplifier laser diode (AL \cite{laser}). By tuning the AL internal diode cavity using its current, the AL's gain can be matched to the injected light frequency. The AL and SL can thereby synchronise frequencies and  also phases. 
Importantly, unlike the broadband gain of a tapered amplifier, the AL current can be tuned to only amplify one narrow-band laser frequency and filter out all others. We can therefore selectively amplify only one fiber-EOM frequency sideband of our SL   \cite{Snadden1997}.
 This allows cheap laser diodes to be used as frequency filtering phase-locked amplifiers without compromising on laser quality.

A measurable parameter and performance metric for OIL is the capture range ($\Delta f_\textrm{c}$) which describes the frequency range over which the AL copies the injected light. For a diode laser, this is given by \cite{Mogensen2018}:
\begin{equation}\label{eq:capturerange}
\Delta f _\textrm{c}=\textrm{FSR}\sqrt{(1+\alpha^2)(R_{\textrm{inj}}-R_\textrm{th})},
\end{equation} where $\textrm{FSR}$ is the free spectral range of the AL  diode cavity (typically $\approx50\,$GHz), $\alpha$ is the linewidth enhancement factor \cite{alpha}, and $R_{\textrm{inj}}=P_{\textrm{inj}}/P_{\textrm{AL}}$ is the ratio of the injected beam power to the AL power.

The current SL in our test setup is a Littman low-power $780.24\,$nm commercial external cavity diode laser (ECDL \cite{vortex}) with a $100\,$GHz mode-hop-free frequency scan, although a Littrow ECDL \cite{Arnold1998} or other laser type \cite{Baillard2006,San2012,Pino2013,Moriya2020,DiGaetano2020,Wang2023,Isichenko2024} could also be used. A small portion of the SL beam was used for low-intensity ($<100\,\upmu\textrm{W/cm}^2$ \cite{Siddons_2008}) spectroscopy \cite{Pizzey2022}, Fig.~\ref{fig:diag} (a,b), through a $74\,\textrm{mm}$ Rb vapour cell. The transmission signal is normalised by dividing with the intensity of a reference beam \cite{Agnew2024}. Narrow-band frequency filtering with a $1.2\,$nm FWHM filter gave a 50-fold reduction to SL amplified spontaneous emission \cite{Agnew2024}. 

\begin{figure}[!t]
    \centering
    \includegraphics[width = .75\columnwidth]{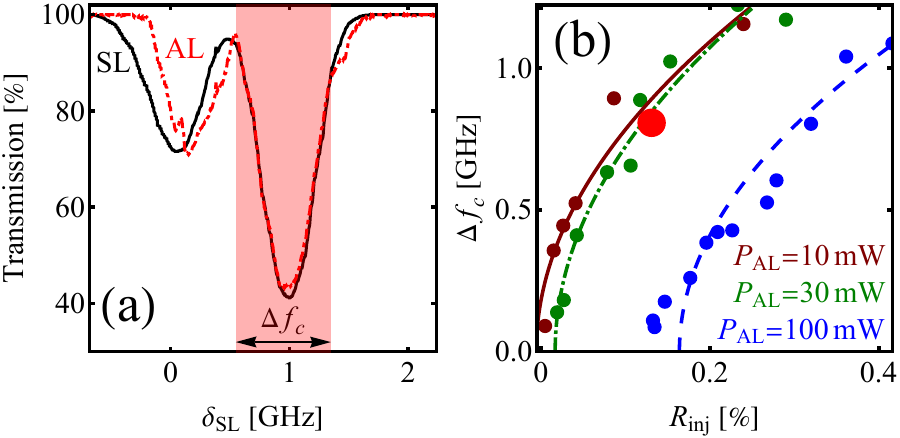}
    \caption{(a) The SL (black) and AL (red, dot-dashed) low-intensity transmission ($T$) spectroscopy signals in a scan of SL detuning $(\delta_\textrm{SL})$ over the two $F=2\leftrightarrow F'$ and $F=3\leftrightarrow F'$ Doppler-broadened $^{87}$Rb and $^{85}$Rb D$_2$ lines, respectively. The highlighted large data point in (b) illustrates our empirical method to extract $\Delta f_\textrm{c}$ from the region where the AL faithfully copies the SL (red shaded region)  (b) Capture range $\Delta f_\textrm{c}$ as a function of injection ratio $R_\textrm{inj}$ for three AL powers $P_{\textrm{AL}} = (10, 30, 100)\,$mW (red, green, blue). The corresponding Eq.~\ref{eq:capturerange} squareroot fits (solid, dash-dotted and dashed) assume a $50\,$GHz FSR, and allow for an $R_\textrm{th}$ offset. \com{The large red marker indicates the dataset shown in (a).}} 
    \label{fig:cap-vs-ir}
\end{figure}

Only small amounts of SL light $\sim1\,$mW are needed for both spectroscopy and injection locking. The latter SL beam was coupled into a single-mode, polarisation-maintaining fibre EOM, and its output (injection beam) was then aligned into the AL cavity via the rejection output port of an optical isolator \cite{reject}. 
To quantify injection locking performance the AL output was  measured in its own low-intensity spectroscopy setup. This enables simple characterisation of the SL vs.\ AL   output using Doppler-broadened dips of the D$_2$ line of the natural isotopes of Rb (Fig.~\ref{fig:cap-vs-ir} (a)).

Correct SL beam alignment into the AL cavity can be optimised by maximising the observed capture range, with Fig.~\ref{fig:cap-vs-ir}~(a) corresponding to SL input and AL output powers of $13\,\upmu\textrm{W}$ and  $10\,\textrm{mW}$, respectively, when using no EOM sidebands. We note the SL beam is fibered and circular, whereas the AL beam profile is elliptical, indicating that better AL-SL mode-matching will enhance injection performance. The behaviour of the capture range, quantified in Eq. (\ref{eq:capturerange}), is demonstrated experimentally in Fig.~\ref{fig:cap-vs-ir}~(b), with better agreement found using a threshold injection $R_\textrm{th}$ ratio that increases with $P_{\textrm{AL}}$ \cite{thresh}.

\begin{figure}[!t]
    \centering
    \includegraphics[width = 1.0\columnwidth]{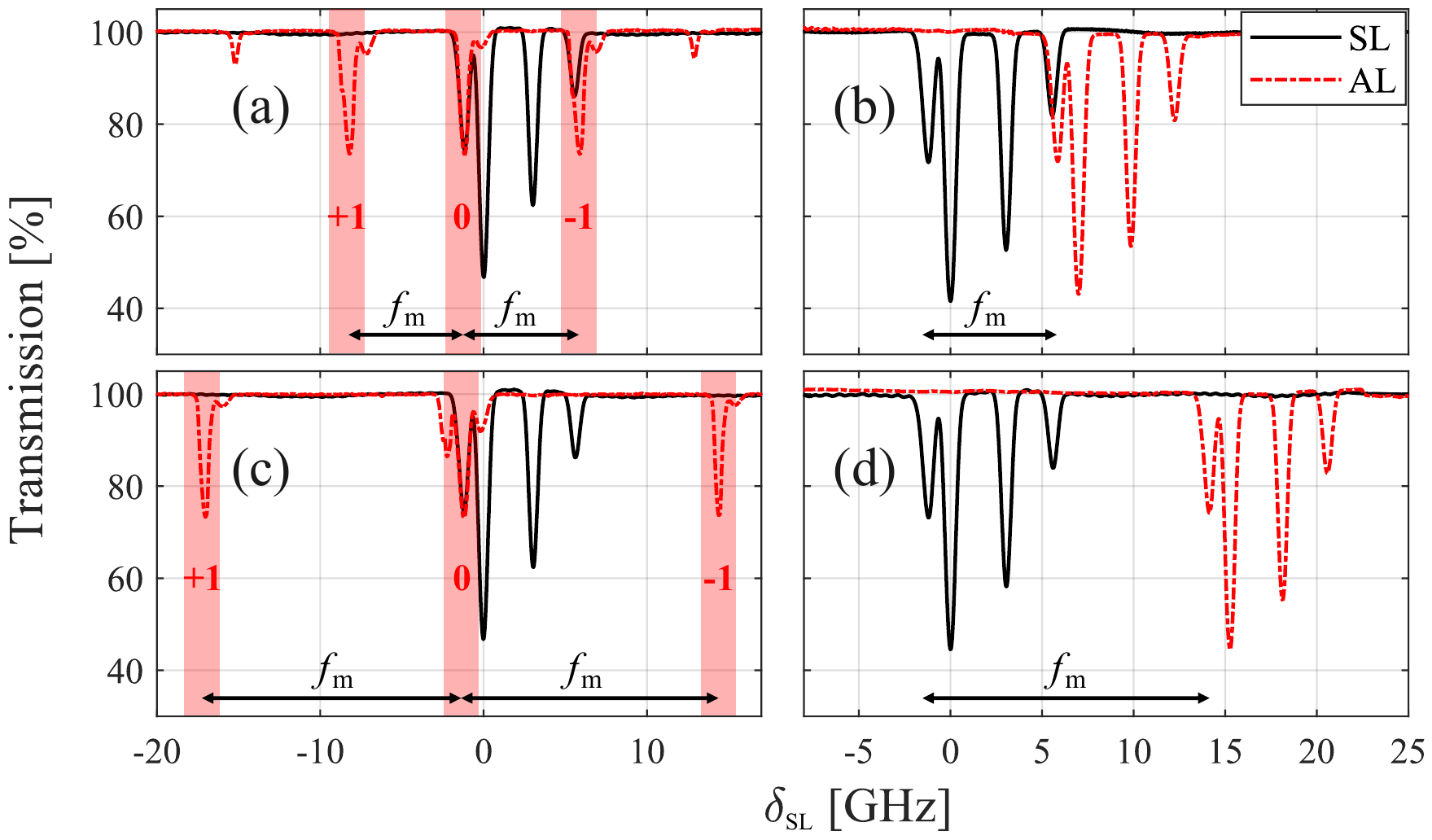}
    \caption{Optical injection locking of the AL with multi-frequency SL injection light at (a) $f_\textrm{m}=6.83\,$GHz and \com{(c) $f_\textrm{m}=15\,$GHz} ($R_{\textrm{inj}}=0.5\,\%$ with  $P_{\textrm{AL}}=81.0\,\textrm{mW}$ and $P_{\textrm{inj}}=0.390\,\textrm{mW}$). The AL current is fixed \com{at a value chosen such that} it only copies the frequency of the $^{87}$Rb $F=2\leftrightarrow F'$ transitions, for different EOM  sidebands, as the SL frequency increases. By using a linear ramp of the AL current, synchronised to the SL frequency scan, the capture range for only the $-1^\textrm{st}$ EOM sideband at (b) $f_\textrm{m}=6.83\,$GHz and \com{(d) $f_\textrm{m}=15.00\,$GHz} is widened.}
    \label{fig:sideband-inj}
\end{figure}

We now consider the AL behaviour when the fiber EOM is activated, with modulation frequency $f_\textrm{m}$ (Fig.~\ref{fig:sideband-inj}). The optical power after the EOM is primarily distributed between the carrier ($0^\textrm{th}$ order) and the $\pm1^\textrm{st}$- order sidebands, with the relative amplitudes determined by the applied RF power and modulation frequency \cite{modsize}. As a result, injection locking to an individual sideband occurs with reduced optical power, leading to a slight reduction in the capture range. 
Despite the lower overall injection power, Fig.~\ref{fig:sideband-inj}~(a) \com{and (c)} show successful injection locking of the AL to the carrier and to each of the first order sidebands for modulation frequencies of $6.83$\,GHz \com{and $15.00$\,GHz}.

In this case, where the SL current is being scanned and the AL current is stationary the injection locking occurs when an SL EOM sideband falls within the AL capture range. This condition is satisfied when the SL detuning is approximately -$f_\textrm{m}$, 0 or +$f_\textrm{m}$, corresponding to the +$1^\textrm{st}$, $0^\textrm{th}$, and -$1^\textrm{st}$ sidebands respectively. In each case the AL amplifies when an SL EOM sideband's frequency is near $\delta_\textrm{SL}\sim0\,$GHz.

\com{A secondary weaker lobe can be observed on the right of each sideband copy in Fig.~\ref{fig:sideband-inj}~(a) and (c). This feature corresponds to the neighboring Rb Doppler resonance, that is only weakly reproduced by the AL.} 
\com{This arises because the capture range is not a sharp step function, as implied by the conceptual diagram in Fig.~\ref{fig:diag} (c), where the AL would produce no copy outside the capture range and a perfect copy within it. Instead, the capture range region is more subtle.}

Applying a continuous linear ramp of the AL current, synchronised to the SL frequency scan, extends the AL's -$1^\textrm{st}$ order single-sideband EOM capture range significantly (Fig.~\ref{fig:sideband-inj}~(b) \com{and (d)} for $f_\textrm{m} = 6.83$\,GHz \com{and $f_\textrm{m} = 15$\,GHz respectively}) \cite{amp}. By changing the AL current one can instead extend the scan range of the $0^\textrm{th}$ or $+1^\textrm{st}$ EOM sideband. \com{The appropriate amplitude scan for the AL was determined by scanning the SL with the EOM switched off, while manually adjusting the current of the AL and recording the current required to reproduce each Doppler feature. This procedure allows the current to be related directly to a capture-range frequency.} 

\begin{figure}[!t]
    \centering
   \includegraphics[width = 0.8\columnwidth]{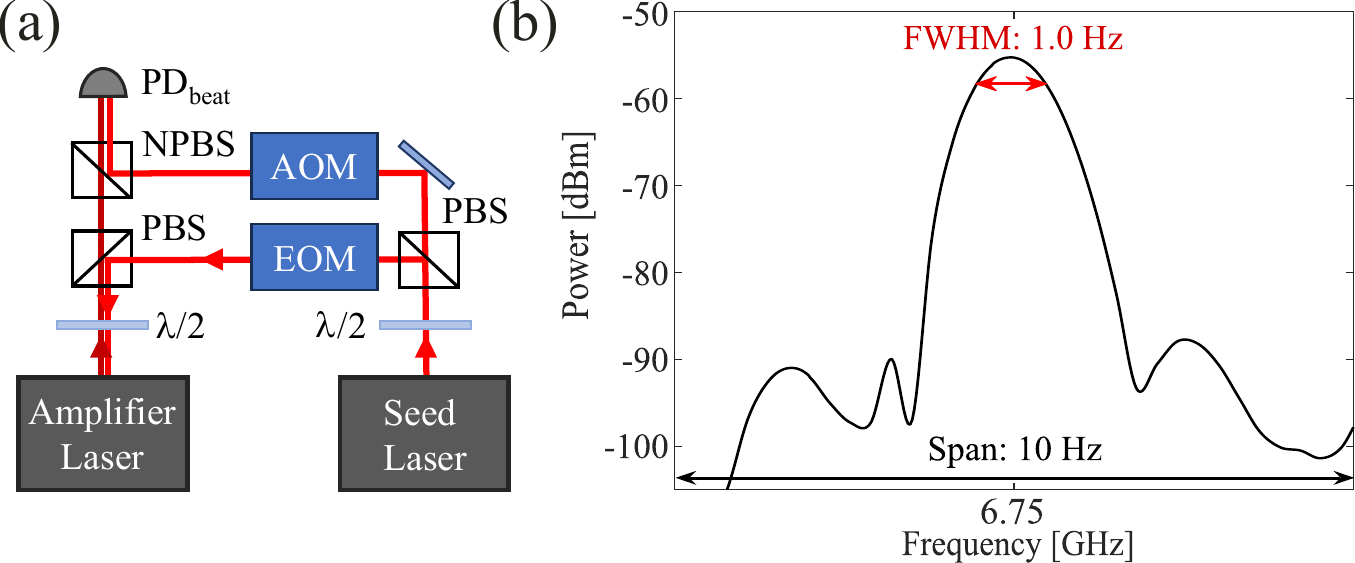}
    \caption{\com{(a) Experimental setup for beat-note measurement. A portion of the SL is sent to the EOM and subsequently injected into the AL. The remaining SL light is sent through an AOM where it is frequency shifted by $0.08\,$GHz. The frequency shifted SL beam and injected AL output are them recombined and detected using a fast PD.} (b) A radio frequency beatnote centred at $6.75\,$GHz between the AL-amplified -$1^\textrm{st}$ order EOM sideband (phase-modulated at $6.83\,$GHz) and the -$0.08\,$GHz-offset SL laser. \com{The $10\,$Hz span is the RF spectrum analyser's minimum, with   RBW and VBW both $1$\,Hz. The beat-note itself does not provide direct evidence of a low phase-error \cite{Snadden1997,Zhang2018,Brodnik2021} but is strongly indicative of one, especially relative to the $>1\,$kHz frequencies of many atomic processes.}}
    \label{fig:beat}
\end{figure}

We demonstrate that the behaviour shown in Fig.~\ref{fig:sideband-inj} extends fairly well to modulation frequencies up to $f_\textrm{m}=15\,$GHz. \com{At low modulation frequencies, the capture range overlaps with multiple EOM sideband orders. As a result, the AL simultaneous copies several modes rather than reproducing a single mode}, setting a practical lower bound on the modulation frequency. Nevertheless, modulation frequencies $<3\,$GHz can still be realised.

While the low-intensity spectroscopy method presented here shows that the AL is a good frequency-offset copy of the SL, it does not in itself prove phase-locking. We therefore also performed an RF beat note measurement, by combining mode-matched co-polarised AL and SL beams on a $(0-25)\,$GHz photodiode \cite{beatPD} \com{as shown in Fig.~\ref{fig:beat}~(a)}. The SL beam used for the beatnote was unmodulated by the EOM, but frequency offset by $-80\,$MHz using an acousto-optic modulator. The resulting beat-note (Fig.~\ref{fig:beat}~\com{(b)}) has a $1\,$Hz full width half maximum linewidth over a $10\,$Hz span -- at the $1\,$Hz frequency resolution limit of the RF spectrum analyser. Moreover, the $-80\,$MHz SL offset allowed us to use the distinct beatnotes to separate and accurately determine the relative amplitudes of any unwanted EOM orders from the AL. We observed a typical AL EOM sideband rejection ratio of $\approx 20\,$dB under the conditions of Fig.~\ref{fig:beat}~\com{(b)}. \com{This result was later cross-checked \cite{Agnew2026} by analysing the AL output with a Fabry-Perot etalon, while the SL was locked, were we found good agreement between the two measurement methods.}

\section{Application note: lasers for Rb gray molasses \label{GM}}

The first observations of sub-Doppler  \cite{Lett1988}, and GM \cite{Boiron1995} cooling schemes now facilitate research into \com{a wide variety of quantum technologies where ultracold or quantum degenerate atoms \cite{Anderson_BEC,Ketterle_BEC} are a pre-requisite, including quantum communication (atomic memories) \cite{Saglamyurek2018,Li2020}, advanced sensing of time \cite{Knappe2004,AbdelHafiz2017,Liu2017,Elgin2019,Elvin2019} and acceleration \cite{Dickerson2013,Rakholia2014,Wu2017,Morel2020,Abend2023}, and quantum computing \cite{Blatt2012,McDonnell2022,Graham2022,Evered2023,Bornet2023}.} 
In particular\com{,} GM cooling \cite{Gabardos:19} has recently yielded striking phase-space density enhancements in several alkali metals including Li \cite{Burchianti14:li6, grier13:li7}, Na \cite{Shi_2018}, K \cite{ang22:39k, Bruce_2017:40k, Rio_Fernandes_2012:40k}, Rb \cite{Huang21:85rb, Rosi2018}, and Cs \cite{Boiron1995, Hsiao18:cs} and it has also been employed in molecules \cite{Truppe2017}, and grating magneto-optical traps \cite{barker2022}.

\begin{figure}[!t]
    \centering
    \begin{minipage}{.32\columnwidth} \centering
    \includegraphics[width = \textwidth]{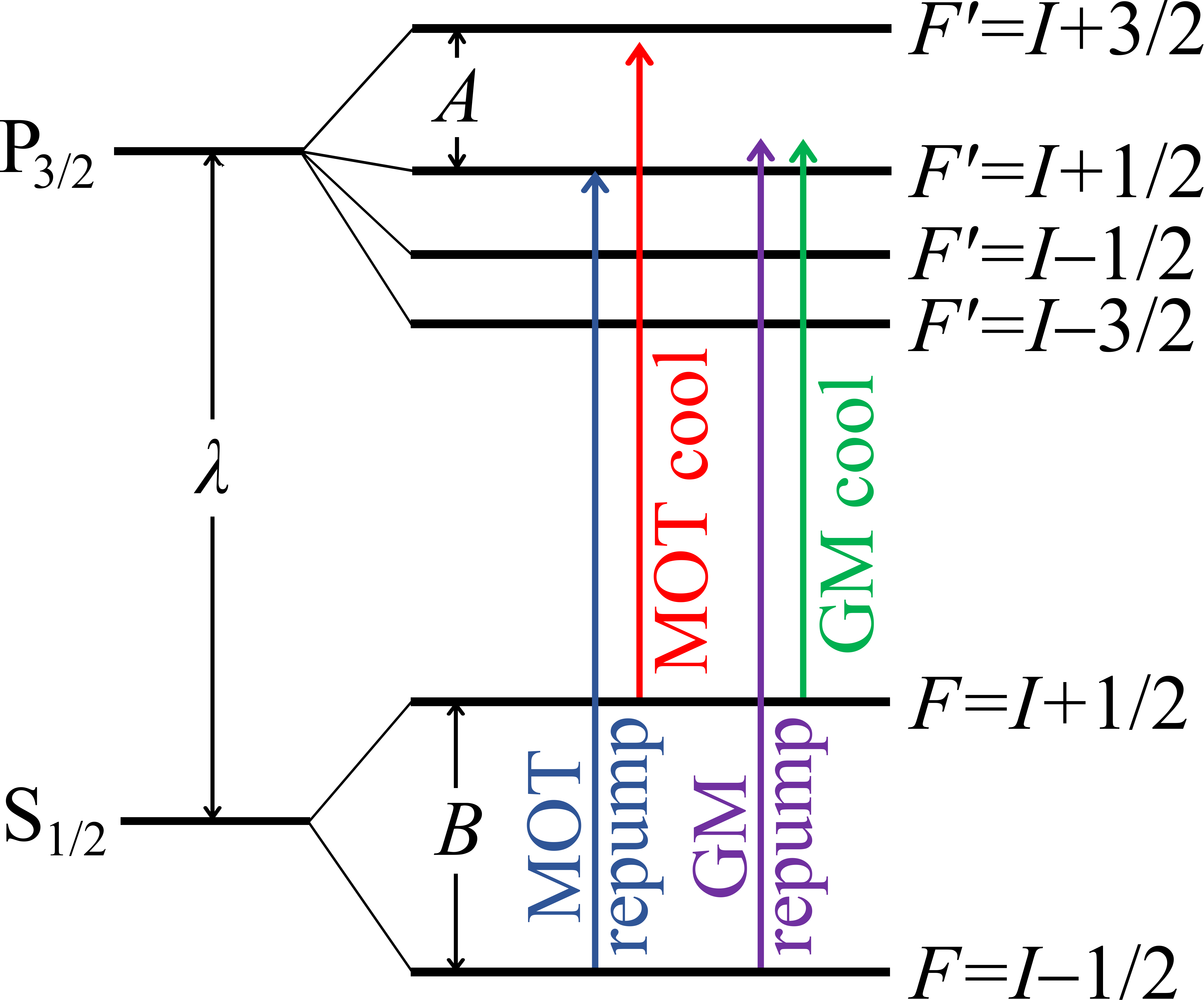}
    \end{minipage}
    \begin{minipage}{.2\columnwidth} \centering
\begin{tabular}{|c|c|c|c|c|}
\hline
$Z$ & $I$ & $\lambda$ & $A$ & $B$ \\ \hline\hline 
Na & 1.5 & 589 & 58 & 1772\\ \hline
$^{40}$K & 4 & 767 & -44 & -1286\\ \hline
$^{85}$Rb & 2.5 & 780 & 121 & 3036\\ \hline
$^{87}$Rb & 1.5 & 780 & 267 & 6835\\ \hline
Cs & 3.5 & 852 & 251 & 9193\\
\hline
\end{tabular}
    \end{minipage}
    \caption{The D$_2$-line hyperfine manifold with MOT and GM transitions for the appropriate alkali metal isotopes $^{23}$Na, $^{40}$K, $^{85}$Rb, $^{87}$Rb and $^{133}$Cs. The nuclear spin is $I$, the wavelength is $\lambda$ in nm, and $A$ and $B$ are hyperfine level splittings in MHz.}
    \label{fig:rb_levs}
\end{figure}

\com{We begin with a brief overview of the laser frequency requirements for D$_2$ cooling  in a generic alkali metal atom with nuclear spin $I$, which usually starts with a magneto-optical trap (MOT \cite{Raab1987,Lee1996,Vangeleyn2009,Vangeleyn2010}). MOT cooling and repumping lasers are the two left transitions in Fig.~\ref{fig:rb_levs}. Cooling is red-detuned typically a few linewidths from the closed $F=I+1/2\rightarrow F'=I+3/2$ transition. Due to Lorentzian atomic absorption with frequency, accidental driving of the $F=I+1/2\rightarrow F'=I+I/2$ occurs over time, from which half of spontaneous decays land in the $F=I-1/2$ ground state, far off-resonant to the cooling laser. A `repump' laser, typically with much lower power, is resonant with the $F=I-1/2\rightarrow F'=I+I/2$ transition and repumps atoms to the $F=I+1/2$ state to return to the cooling cycle. The repump laser is not necessarily phase-locked with the cooling laser, and they often have $1\,$MHz-level  uncorrelated relative frequency noise.

Further sub-Doppler cooling in optical molasses can be achieved by cancelling all magnetic fields and further red-detuning the cooling laser \cite{Lett1988}. After a few milliseconds of optical molasses, gray molasses can be applied using the two laser transitions on the right of the Fig.~\ref{fig:rb_levs} level diagram. As the transitions share a connecting $F'=I+1/2$ excited state, this $\Lambda$-transition can support coherent dynamics and dark states -- so long as the laser frequencies are phase-stable with respect to each other, as clearly demonstrated in Ref.~\cite{Rosi2018}. The gray molasses repump laser also plays a more important role in the cooling process and has to have much higher power \cite{Rosi2018}.}

\com{To most easily obtain the two laser frequencies for all cooling stages -- the MOT, optical molasses and gray molasses -- an $\approx 100\,$mW SL `repump' laser can be locked to the $F=I-1/2\leftrightarrow F'=I+1/2$ transition (cf.~Fig.~\ref{fig:diag} (c) (1)), and a small portion of modulated light from this SL creates a `cooling' AL that can be red-detuned (cf.~Fig.~\ref{fig:diag} (c) (3))} from approximately $B-A$ to $B$ (Fig.~\ref{fig:rb_levs}), using the EOM modulation frequency. To be specific, for $^{87}$Rb the D$_2$ SL laser needs a sub-Doppler lock to the $F=1\leftrightarrow F'=2$ transition (e.g.\ using hyperfine pumping spectroscopy \cite{Smith2004}), and we wish the AL to amplify only the red sideband of the fiber EOM in the range of 6.58 to 6.83$\,\textrm{GHz}$. This produces a phase-locked cooling transition beam which can be optically combined with the repump beam for all stages of cooling.

\com{We explicitly consider the laser sources required for GM cooling of rubidium here.}  Existing GM cooling studies in $^{85}$Rb and $^{87}$Rb report using laser powers of order $100\,\textrm{mW}$ \cite{Rosi2018,Huang21:85rb}. Although stable OIL with a small capture range has been demonstrated in diode lasers with low injection ratios $\approx 10^{-5}$ \cite{Kakarla:18}, it is beneficial to have a higher injection power as a larger capture range provides reduced intensity noise in the AL \cite{Liu2013}. To achieve $100\,\textrm{mW}$ powers over the $0.5\,\textrm{GHz}$ capture range needed for Rb gray molasses, sideband injection powers $<1\,\textrm{mW}$ suffice (Fig.~\ref{fig:cap-vs-ir}~(a)).

For GM using the D$_2$ atomic line, e.g.\ $^{40}$K \cite{Bruce_2017:40k}, $^{87}$Rb \cite{Rosi2018} and Cs \cite{Hsiao18:cs}, 
or even the Li D$_1$ line \cite{Burchianti14:li6, grier13:li7}, our laser system tunability means that the \textit{same} laser system can be used for regular magneto-optical trapping and optical molasses, prior to GM cooling, obviating the need for an additional GM laser system which is sometimes used \cite{Boiron1995}. 
While applications are not limited to this species, our D$_2$ line laser system is tested with GM cooling of Rb in mind. For parameters relevant for other alkali metal species see the general level diagram and table in Fig.~\ref{fig:rb_levs}.

\section{Conclusion and outlook}

In conclusion, we have developed a simple, high-power, relative-frequency-tunable phase-locked two-laser system that is ideal for magneto-optical trapping, optical molasses and gray molasses using the same D$_2$ setup. Moreover, the design can also be useful in clock, interferometry, quantum computing and alkali metal Doppler thermometry \cite{Truong2011,Agnew2024} experiments.
The architecture could also conceivably be used for future on-chip compact phase-locked laser systems for quantum technologies \cite{McGilligan2022}. The phase-locking quality could be characterised further using the approach in \cite{Snadden1997}, and the system could be improved in future by achieving better mode-matching between the SL and AL beam shapes, and using a higher-power AL laser diode \cite{Daffurn2021}. For higher Watt-level powers a tapered amplifier could be used on the AL, after the AL diode removes unwanted EOM sidebands.

\section*{Availability of supporting data} Data underlying the results presented in this paper are available in Ref.~[TBD].

\section*{Competing interests} The authors declare no conflicts of interest.

\section*{Funding} 
Engineering and Physical Sciences Research Council EP/T001046/1 and EP/Z533166/1; National Physical Laboratory iCASE studentship EP/X525017/1.

\section*{Acknowledgments} We thank the wider EQOP quantum technology team for general support. Special  thanks to Jon Pritchard and Erling Riis for proof-reading and improving the manuscript, and Graham Machin, Paul Griffin and Sonja Franke-Arnold for valuable discussions. For the purpose of open access, the
authors have applied a Creative Commons Attribution (CC BY)
licence to any Author Accepted Manuscript (AAM) version
arising from this submission.

\bibliography{references.bib}

\end{document}